\documentclass[symmetry,article,submit,oneauthor,latex,dvi2pdf,12pt,a4paper]{mdpi} 
\lastpage{x}
\doinum{10.3390/------}
\pubvolume{xx}
\pubyear{2012}
\history{Received: xx / Accepted: xx / Published: xx}

\Title{Fermi surface reconstruction due to hidden 
rotating antiferromagnetism in n and p-type high-$T_C$ cuprates}

\Author{Mohamed Azzouz}

\address[1]{
Department of Physics, Laurentian University,
Ramsey Lake Road. Sudbury, 
Ontario, Canada P3E 2C6}

\corres{mazzouz@laurentian.ca, Tel. 1-705 675 1151, Ext. 2224
and Fax 1-705 675 4868.}

\abstract{The Fermi surface calculated within the rotating 
antiferromagnetism theory
undergoes a topological change when doping changes from p-type to n-type,
in qualitative agreement with experimental data for 
n-type cuprate Nd$_{2-x}$Ce$_x$CuO$_4$ and 
p-type La$_{2-x}$Sr$_x$CuO$_4$. 
Also, the reconstruction of the Fermi surface, 
observed experimentally close to optimal doing in p-type cuprates,
and slightly higher than optimal doping in the overdoped regime
for this n-type high-$T_C$ cuprate, is well accounted 
for in this theory. This reconstruction is 
a consequence of the quantum criticality 
caused by the disappearance of rotating antiferromagnetism. 
The present results 
are in qualitative agreement with recently observed
quantum oscillations in some high-$T_C$ cuprates. This paper 
presents new results about the application of the rotating 
antiferromagnetism theory to the study of the electronic structure 
for n-type materials.}

\keyword{Rotating antiferromagnetism; High-T$_C$ cuprates; 
Hidden order; Symmetry breaking; Fermi surface reconstruction}

\begin{document}

\section{Introduction}

The topology and doping dependence of the Fermi surface (FS) 
of high-temperature 
superconductors (HTSC)
are currently highly debated.
Some observations from angle-resolved-photoemission spectroscopy (ARPES)
experiments do not seem to see any FS 
reconstruction, but
data collected from magnetoresistance measurements 
characterized by Shubnikov-de Haas (SdH) oscillations
indicate that the FS undergoes 
a topology change due to 
some sort of symmetry breaking. Since no 
long range order has been observed so far 
in underdoped HTSCs we proposed earlier that
the FS reconstruction is caused by the hidden 
rotating antiferromagnetic order.

In the present work we support this proposal
by new results for n-type cuprates and argue in  
favor of the FS of HTSCs undergoing
topology reconstruction at specific doping levels
in the framework of rotating antiferromagnetism
theory (RAFT) \cite{azzouz2003}.
We compare the evolution of the FS with doping in p-type 
and n-type HTSCs obtained in this theory, and discuss it
in connection mainly with
available experimental data for n-type material
Nd$_{2-x}$Ce$_x$CuO$_4$ and p-type one La$_{2-x}$Sr$_x$CuO$_4$.
It is found that 
the change in the topology of the FS as one goes 
from the p-type cuprate to n-type material 
is well accounted for in RAFT.
In the low-doping limit (underdoped regime)
RAFT yields a small almost square FS {\it centered} around
$(\pi,0)$ points for n-type 
Nd$_{2-x}$Ce$_x$CuO$_4$ in qualitative 
agreement with 
SdH oscillations, which indicate the existence 
of a FS in the form of small pockets \cite{helm2009}.
A careful look at the data of Armitage 
{\it et al.} \cite{armitage2002}
in Fig. 3 of their work reveals a trend 
qualitatively consistent with our findings
for n-type material Nd$_{2-x}$Ce$_x$CuO$_4$ regarding the
evolution of spectral weight 
away from $(\pi,0)$ and the formation of a larger FS as doping
increases.
A FS in the form of stretched elliptic pockets nearby
the $(\pi/2,\pi/2)$ points is however likely 
for p-type La$_{2-x}$Sr$_x$CuO$_4$. Indeed, Fig. 5 of the ARPES 
work by
Yoshida {\it et al.} \cite{yoshida2006} shows nicely
the evolution of the FS with doping from what we interpret as
stretched small pockets
in the underdoped regime to large contours in the overdoped regime.
RAFT reproduces qualitatively well 
the FS evolution with doping for this p-type material.  
Note that p-type cuprates were examined 
using RAFT in Ref. \cite{azzouz2010}. However this is the first work
based on RAFT, which deals with 
the electronic structure in an n-type cuprate.
In RAFT, for both p-type and n-type materials,
the critical value of doping where FS reconstruction 
occurs is given by the 
value where rotating antiferromagnetism vanishes. 
In p-type materials,
this value coincides practically with optimal doping, 
but in n-type case, it
occurs in the overdoped regime beyond 
optimal doping for superconductivity (SC).

This paper is organized as follows.
In Section \ref{sec2}, RAFT is extensively reviewed. 
In Section \ref{sec3} the rotating antiferromagnetic 
and superconducting parameters are calculated as a 
function of doping and temperature.
In Section \ref{sec4}, the doping dependence of the 
electronic structure is calculated and compared 
to experimental data. Energy spectra versus wavevector
are calculated for several doping levels, and
the FS is calculated 
using the occupation
probability for doping levels in n-type and p-type cases.
Conclusions and a discussion of existing experimental data 
are given in Section \ref{conclusions}.



\section{Review of Rotating antiferromagnetism theory}

\label{sec2}

\subsection{Normal state}

\label{sec2.1}
We first focus on the normal 
(non superconducting) state where we review the derivation
of rotating antiferromagnetism (RAF). In section 
\ref{RAF and SC} we will review the interplay between 
SC and RAF. 
Consider here the $t$-$t'$ Hubbard model in two dimensions:
\begin{eqnarray}
H&=&
-t\sum_{\langle i,j\rangle\sigma}c^{\dag}_{i,\sigma}c_{j,\sigma} 
-t'\sum_{\langle\langle i,j\rangle\rangle\sigma}c^{\dag}_{i,\sigma}c_{j,\sigma}
+ {\rm h.c.} 
\cr
&&-\mu\sum_{i,\sigma}n_{i,\sigma} + U\sum_in_{i\uparrow}
n_{i\downarrow},
\label{hamiltonian1}
\end{eqnarray}
where $\langle i,j\rangle$ and $\langle\langle i,j\rangle\rangle$
designate nearest and second-nearest neighboring sites,
respectively, and $t$ and $t'$ are 
electron hopping energies to nearest 
and second-nearest neighbors, 
respectively. Note that hopping to further neighbors
was also considered \cite{azzouz2010} for more 
accurate comparison with experiment. 
The interacting term in Hamiltonian 
(\ref{hamiltonian1}) has been decoupled using 
\begin{equation}
Q_i={\langle c_{i,\uparrow}c^{\dag}_{i,\downarrow}\rangle}
=-\langle S_i^-\rangle\equiv|Q|e^{i\phi_i}, 
\label{RAF order parameter}
\end{equation}
and mean-field 
theory \cite{azzouz2003,azzouz2003b,azzouz2004,azzouz2005,azzouz2010,
azzouz2010b} 
was recently combined with 
the Heisenberg equation \cite{azzouz2012} in order to 
calculate the phase of the this order 
parameter. To use the 
Heisenberg equation the interacting 
term $Un_{i\uparrow}n_{i\downarrow}$ was rewritten in terms of
the spin ladder operators in the following way.
In second quantization, where
$S_i^+=c^{\dag}_{i,\uparrow}c_{i,\downarrow}$, 
the onsite Coulomb repulsion $Un_{i\uparrow}
n_{i\downarrow}$ was on one hand written as 
$Un_{i\uparrow}n_{i\downarrow}=Un_{i\uparrow} - US_{i}^+S_i^-$
and on the other hand as
$Un_{i\uparrow}n_{i\downarrow}=Un_{i\downarrow} - US_{i}^-S_i^+$.
Summing then dividing by 2 gave the symmetric expression
$Un_{i\uparrow}n_{i\downarrow}=\frac{U}{2}(n_{i\uparrow} 
+ n_{i\downarrow}) -\frac{U}{2}(S_{i}^+S_i^- +
S_{i}^-S_i^+)$ \cite{azzouz2012}.
The terms 
$S_{i}^+S_i^-$ and  $S_{i}^-S_i^+$, which are responsible for
onsite spin-flip excitations, contribute by lowering 
energy for the sites that are partially occupied by the same density
of spin up and down electrons.
We decoupled 
this term in mean-field theory using $\langle S_i^-\rangle\equiv 
\langle c^{\dag}_{i,\downarrow}c_{i,\uparrow}\rangle$, 
which leads to a 
collective behavior for the spin-flips, 
and recovered the results
obtained earlier in RAFT 
\cite{azzouz2003,azzouz2003b,azzouz2004,azzouz2005,azzouz2010,azzouz2010b}.
In this state, a spin flip process at site $i$ 
is simultaneously accompanied by another one at another site $j$; 
the occurrence of spin flips becomes synchronized 
below a transition temperature, which was identified with the 
pseudogap (PG) temperature. In Section \ref{sec2.3} below, an interpretation
of RAF from a classical point of view will be given.

The parameter 
$Q_i$ in (\ref{RAF order parameter}) is thus used to carry on
a mean-field decoupling of the $t$-$t'$ Hubbard model 
(\ref{hamiltonian1}). Consideration of the 
ansatz where $\phi_i-\phi_j=\pi$, 
with $i$ and $j$ labeling any two adjacent
lattice sites, and letting the phase $\phi_i\equiv\phi$
be site independent but assuming any value in $[0,2\pi]$ led to
the following normal state Hamiltonian in 
RAFT \cite{azzouz2003,azzouz2003b,azzouz2004}
\begin{equation}
H\approx\sum_{{\bf k}\in RBZ}\Psi^{\dag}_{\bf k}{\cal H}\Psi_{\bf k}
+NUQ^2-NUn^2,
\label{raft hamiltonian}
\end{equation}
where $N$ is the number of sites, and
$n=\langle n_{i,\sigma}\rangle$ is the expectation value 
of the number operator. 
Because of antiferromagnetic
correlations a bipartite lattice with sublattices
$A$ and $B$ is considered, even though 
no long-range static antiferromagnetic order is taken into account.
Note that RAFT is only valid away from half-filling
where this long-range order occurs.
The summation runs over the reduced (magnetic)
Brillouin zone (RBZ). The Nambu spinor is
$
\Psi^{\dag}_{\bf k}=(c^{A\dag}_{{\bf k}\uparrow}\ c^{B\dag}_{{\bf k}
\uparrow}\ c^{A\dag}_{{\bf k}\downarrow}\ c^{B\dag}_{{\bf k}\downarrow})
$, and the Hamiltonian matrix is 
\[
{\cal H}=
\left( 
\begin{array}{ccccc}
&-\mu'  &\epsilon&Qe^{i\phi}&0 \\ 
&\epsilon & -\mu'&0 & -Qe^{i\phi} \\
&Qe^{-i\phi}&0 &-\mu' &\epsilon \\
&0 & -Qe^{-i\phi}&\epsilon &-\mu'\\
\end{array}
\right),
\]
yielding the energy spectra
\begin{equation}
E_{\pm}({\bf k})=-\mu'({\bf k})\pm E_q({\bf k}),
\label{normal state spectra}
\end{equation}
where
%
%
$
\mu'({\bf k})=\mu -Un +4t'\cos k_x \cos k_y 
$,
%
%
$E_q({\bf k})=\sqrt{\epsilon^2({\bf k})+(UQ)^2}$, and 
%
%
$
\epsilon({\bf k})=-2t(\cos k_x+\cos k_y)
$.
Using the fact that the energy spectra $E_\pm({\bf k})$ 
do not depend on phase $\phi$,
the matrix $\cal H$ is transformed 
to one that does not depend
on $\phi$ using the spin-dependent 
gauge transformation
%
%
$
c_{i,\uparrow}\to e^{i\phi/2}c_{i,\uparrow}$
and
$
c_{i,\downarrow}\to e^{-i\phi/2}c_{i,\downarrow}$.
%
%
%
This transformation is equivalent to performing 
a rotation by angle $-\phi$ about the 
$z$ axis for the $x$ and $y$ components of the spin operator 
according to:
\[
\left( 
\begin{array}{cc}
&S_i^x\\ 
&S_i^y
\end{array}
\right)
\to
\left( 
\begin{array}{cc}
\cos\phi  &\sin\phi \\ 
-\sin\phi  &\cos\phi
\end{array}
\right)
\left( 
\begin{array}{cc}
&S_i^x\\ 
&S_i^y
\end{array}
\right).
\]
Note that the thermal 
averages of $S_i^x$ and $S_i^y$ are given by
\begin{eqnarray}
\frac{\langle S_i^x\rangle}{\hbar} &=& Q\cos\phi,\  
\frac{\langle S_i^y\rangle}{\hbar} = -Q\sin\phi,\ \ 
i \in A, \ {\rm or}\cr
\frac{\langle S_i^x\rangle}{\hbar} &=& -Q\cos\phi,\ 
\frac{\langle S_i^y\rangle}{\hbar} = Q\sin\phi,\ \ 
 i \in B,
\label{mag config}
\end{eqnarray}
and $\langle S_i^z\rangle=0$ for $i$ in both sublattices.
Because the phase $\phi$ assumes 
any value
between $0$ and $2\pi$, rotational symmetry will not look broken
for times greater than the period of rotation as 
we will explain below, when we review the calculation of 
the time dependence of the phase.
However if the typical time scale of a probe is much smaller than 
this period symmetry may appear broken. 

The magnitude $Q$ and electron occupation thermal average $n$
are calculated by minimizing the phase-independent
mean-field free energy. The following mean-field 
equations were obtained in the normal state \cite{azzouz2003,azzouz2003b,azzouz2004} 
\begin{eqnarray}
n&=&\frac{1}{2N}\sum_{\bf k}\{n_F[E_+({\bf k})] + n_F[E_-({\bf k})]\}\cr
Q &=&\frac{U}{2N}\sum_{\bf k}\frac{n_F[E_-({\bf k})] 
- n_F[E_+({\bf k})]}{E_q({\bf k})}.
\label{free energy}
\end{eqnarray}
\subsection{Calculation of the time dependence of the phase}
\label{sec2.2}

The nature of RAF has recently been completely understood
after the phase $\phi$ of its order parameter was calculated as
a function of time \cite{azzouz2012}. Here we summarize how this was done.
The Heisenberg equation 
$\frac{dS_j^+}{d\tau}=\frac{1}{i\hbar}[S_j^+,H]$ was calculated
in the limit where 
electron hopping is neglected 
in comparison to $\frac{U}{2}(S_j^+S_j^- + S_j^-S_j^+)$. 
The values considered in RAFT for onsite Coulomb repulsion are in the range
$U\sim 3t$-$5t$; 
this is an intermediate coupling regime where $U>t$ 
but smaller than the bandwidth $\sim8t$ when $t'\ll t$.
Neglecting the effect of electron 
hopping energies in the Heisenberg equation 
can be justified on the ground that
spin dynamics is faster than
charge dynamics. An onsite 
spin flip fluctuation needs a time $\tau\sim \hbar/U$ to be realized, while 
a fluctuation caused by a charge hopping between adjacent sites 
takes a longer time $\tau\sim\hbar/t$, ($U>t$). In
the Heisenberg equation the bare original interaction was used
instead of RAFT's Hamiltonian (\ref{raft hamiltonian})
in order to treat as best as possible quantum fluctuations.
In this approximation, the following time equation 
was obtained \cite{azzouz2012}
\begin{equation}
\frac{dS_j^+}{d\tau}\approx i\frac{U}{\hbar}S_j^+,\ \ (\tau\ {\rm is\ time}),
\label{heisenberg eq}
\end{equation}
in the intermediate regime where
spin dynamics is not governed by the Heisenberg exchange 
coupling $ 4t^2/U$. Note that the latter is suitable in the 
strong coupling limit ($U/t\gg1$) for the Hubbard model, whereas 
RAFT is valid in the intermediate coupling regime.
Integration over time $\tau$ in 
Eq. (\ref{heisenberg eq}) 
gives for the thermal average
\begin{eqnarray}
\langle S_j^+(\tau)\rangle\approx \langle S_j^+(0)\rangle e^{iU\tau/\hbar}.
\label{sol heisenberg eq}
\end{eqnarray}
The phase can thus be written as $\phi=U\tau/\hbar$ modulo $2\pi$
when $\langle S_j^+(0) \rangle$ is identified with  
$|\langle S_j^+(\tau)\rangle|$, 
($-|\langle S_j^+(\tau)\rangle|$), for sublattice $A$, ($B$),
and $e^{i\phi}$ with $e^{iU\tau/\hbar}$. Using this result,
the magnetic configuration (\ref{mag config}) is rewritten as 
follows  
$
\langle S_i^x\rangle/\hbar = Q\cos(\omega_{sf} \tau)$, 
$\langle S_i^y\rangle/\hbar = -Q\sin(\omega_{sf} \tau)$
for $i$ in sublattice $A$ or
$\langle S_i^x\rangle/\hbar = -Q\cos(\omega_{sf} \tau)$, 
$\langle S_i^y\rangle/\hbar = Q\sin(\omega_{sf} \tau)$
for $i$ in sublattice $B$, and
$\langle S_i^z\rangle = 0$ 
for $i$ in sublattice $A$ or $B$.
These thermal averages describe a rotational motion for the 
spin components with angular frequency
$\omega_{sf}=U/\hbar$, and period 
$T_{sf}=2\pi\hbar/U$ is the time required 
to perform a spin-flip process, or the time needed for 
the rotating order parameter $\langle S_i^{x(y)} \rangle$
to complete a $2\pi$-revolution in a classical picture.

\subsection{Interpretation of rotating antiferromagnetism}
\label{sec2.3}

The above derivation of RAF was supported by the following argument,
which shows that rotating magnetism (ferro or antiferro) is physically
sound and can therefore be realized in a real 
system independently of a model.
Consider the much simpler case of a 
single spin precessing in a magnetic field 
$B$ along the $z$-axis, 
with the initial spin state given by
$| S_x,+\rangle = {\frac{1}{\sqrt 2}}(|\uparrow\rangle + |\downarrow\rangle)$.
Initially, this spin points in the positive $x$-direction.
The time-dependent
expectation values of this spin's components are
$\langle S^x\rangle=\frac{\hbar}{2}\cos(\omega t)$,
$\langle S^y\rangle=\frac{\hbar}{2}\sin(\omega t)$, and 
$\langle S^z\rangle=0$, with $\omega=\frac{|e|B}{m_ec}$. 
$e$ and $m_e$ are the charge and mass of the electron, 
and $c$ is the speed of light. 
The $x$ and $y$ components are therefore confined to rotate about the $z$-axis
in the $xy$ plane with Larmor angular frequency $\omega$.
A rotating ferromagnetic 
state can be realized by 
placing $N$ such states with the same 
frequency on a lattice made of $N$ sites.
For a rotating antiferromagnetic state, opposite initial 
states ($\pm| S_x,+\rangle$)
where spins point in opposite directions on the $x$-axis 
are placed on any two adjacent sites of a lattice. 
To relate RAF to spin flip processes, it is noted that 
$\langle S^\pm\rangle=\langle S^x\rangle \pm 
i\langle S^y\rangle=\frac{\hbar}{2}e^{\pm i\omega t}$ 
in this example. 
In a given model, a coupling 
is necessary for providing the building bloc for RAF, which 
is a spin precessing about an effective magnetic field 
(with no local magnetization) 
for each lattice site and the anti-alignment
of the adjacent rotating moments. The RAF state constructed in this way 
shows a hidden order that can be realized even at finite 
temperature without violating the Mermin-Wagner theorem
\cite{mermin1966}.

The above simple case allowed us to 
interpret RAF as 
a state where spins precess collectively 
in a synchronized way
in the spins' $xy$ plane
around a staggered effective 
magnetic field $B=m_ecU/\hbar|e|$ 
generated by onsite Coulomb repulsion. 
$\hbar/2$ in $\langle S^\pm\rangle=\frac{\hbar}{2}e^{\pm i\omega t}$
is replaced by the magnitude of the
RAF order parameter $Q\hbar$, which
assumes
values smaller than $\hbar/2$ due to thermal fluctuations.
In comparison to ordinary spin waves in an antiferromagnet, 
RAF's state was interpreted
as a single ${\bf q}=(\pi,\pi)$ spin wave  
occurring as a consequence of zero staggered static magnetization.  
The spin-wave theory does not however apply 
for our system (where $\langle S_i^z\rangle=0$), since
this theory has to be built on top of a N\'eel background
with finite $\langle S_i^z\rangle$.
Also, in comparison to ordinary antiferromagnetic spin-density order,
RAF is characterized by a local magnetization that is not static
because of the time dependence of the phase of the magnetization.
It is thus clear that RAF will have all the typical effects of
spin-density order on the evolution of the electronic 
structure with doping, but is expected to go undetected 
for experimental probes like neutrons due to the time 
dependence of the phase. We predicted \cite{azzouz2012} that rotational 
symmetry will not look broken 
for experimental probes that are characterized 
by a time scale greater than the period of rotation 
$T_{sf}=2\pi\hbar/U$ of the rotating 
order parameter of RAF. For such probes, averaging 
over times longer than the period will
not allow for the observation of RAF. 
In RAFT, electron hopping energy $t$ is taken to be 
0.1 eV in fitting data. Taking $U=3t=0.3$ eV
gives $T_{sf}\approx 10^{-14}$ s. For neutrons for example
the typical time would be the time spent by 
a given neutron in the immediate vicinity of a given spin during the 
scattering process.
If this time is greater than the period $T_{sf}$ then neutrons will 
not detect RAF. If the time 
spent by the neutron in the vicinity of the spin
is smaller then there is a chance RAF will be detected.
Note that smaller times means higher energies for neutrons. 
This is an issue that is still under investigation and will be reported on
in the future.

\subsection{Interplay between RAF and SC}

\label{RAF and SC}

In RAFT, d-wave SC was introduced phenomenologically
using an attractive coupling  
between electrons on adjacent sites. 
The term $- V\sum_{\langle i,j\rangle}
n_{i,\uparrow}n_{j,\downarrow}$
is now added 
to Hamiltonian (\ref{hamiltonian1}),
%
%
%
%
%
%
and is decoupled using 
$D_{\langle i,j\rangle}
=\langle c_{i,\downarrow}c_{j,\uparrow}\rangle$. 
To get a d-wave gap we set
$D_{\langle i,j\rangle}=D_0$ along the $x$-direction and 
$D_{\langle i,j\rangle}=-D_0$ along the $y$-direction \cite{azzouz2003,azzouz2003b}.

When both SC and RAF
orders are taken into account,
the mean-field Hamiltonian is written in terms
of an eight-component spinor given by 
\begin{eqnarray}
\Psi_{\bf k}^{\dag}
=({c^{A\dag}_{-{\bf k}\uparrow}} {c^{B\dag}_{-{\bf k}\uparrow}}
c^{A}_{{\bf k}\downarrow} c^{B}_{{\bf k}\downarrow}
c^{A}_{{\bf k}\uparrow} c^{B}_{{\bf k}\uparrow}
{c^{A\dag}_{-{\bf k}\downarrow}} {c^{B\dag}_{-{\bf k}\downarrow}}),
\label{vector}
\end{eqnarray}
and  assumes the expression \cite{azzouz2003,azzouz2003b}
\begin{eqnarray}
H=&&\sum_{\bf k_<}\Psi_{\bf k}^{\dag}{\cal H}\Psi_{\bf k}
+ UNQ^2 + UNm^2 \cr &&+ 4VND_0^2 - UNn^2 - \mu N,
\label{hamstaggered}
\end{eqnarray}
where ${\cal H}$ is an $8\times8$ matrix:
\[
{\cal H}=
\left(
\begin{array}{ccc}
{\cal H}' & {\cal U}_Q \\ 
-{\cal U}_Q & -{\cal H}'\\ 
\end{array}
\right) 
\]
with ${\cal H}'$ and ${\cal U}_Q$, two $4\times4$ matrices, given
by
\[
{\cal H}'=
\left(
\begin{array}{ccccc}
-\mu'({\bf k}) & \epsilon({\bf k}) & 0 & D({\bf k}) \\ 
\epsilon({\bf k})& -\mu'({\bf k}) & D({\bf k}) & 0  \\
0 & D({\bf k})  &\mu'({\bf k}) & -\epsilon({\bf k}) \\
D({\bf k}) & 0 & -\epsilon({\bf k}) & \mu'({\bf k}) \\
\end{array}
\right)
\]
and
\[
{\cal U}_Q=
\left(
\begin{array}{ccccc}
0 & 0 & QU & 0 \\ 
 0 & 0 & 0 & -QU \\
-QU & 0 & 0 & 0 \\
 0 & QU & 0 & 0  \\  
\end{array}
\right).
\]
The ${\bf k}$-dependent superconducting gap 
is $D({\bf k})=2VD_0(\cos k_x - \cos k_y)$ where
$D_0=\langle c^A_{2i,j,\downarrow}c^B_{2i\pm1,j\uparrow}\rangle=
-\langle c^A_{i,2j,\downarrow}c^B_{i,2j\pm1,\uparrow}\rangle$
involves two adjacent sites on different sublattices. 
Triplet SC \cite{kyung2000} is ruled out by choosing
$\langle c^A_{2i,j,\downarrow}c^B_{2i\pm1,j\uparrow}\rangle=-
\langle c^A_{2i,j\uparrow}c^B_{2i\pm1,j,\downarrow}\rangle$; so
only spin-singlet SC, which is relevant to HTSCs, is considered in RAFT.
The way decoupling is done using creation and annihilation operators
rather than the spin-singlet and triplet superconducting operators,
which are a combination of products of the $c$'s and $c^{\dag}$'s
allowed us to avoid the generation of triplet SC if it is 
not present initially \cite{kyung2000}.

In Hamiltonian (\ref{hamstaggered}),
$\epsilon({\bf k})=-2t(\cos k_x+\cos k_y)$,
and
%
%
$\mu'({\bf k})=\mu-Un+4t'\cos k_x\cos k_y$
%
%
have the same expressions as in the absence of SC.
In equation (\ref{hamstaggered}), 
the summation over {\bf k} takes
into account the doubling of the Brillouin zone and the fact that 
summation is now over ${\bf k}$ and ${-\bf k}$. 
The size of the mean-field Hamiltonian matrix ${\cal H}$
is twice as large as that in the density $d$-wave (DDW) 
approach \cite{chakravarty}, 
proposed for the PG behavior, 
or in other approaches that deal 
with the interplay between antiferromagnetism and SC \cite{kyung2000}.

The energy spectra obtained by diagonalizing the matrix ${\cal H}$ are
$\pm E_1({\bf k})$ and $\pm E_2({\bf k})$ with
\begin{eqnarray}
E_\nu({\bf k})={\sqrt{[\mu'({\bf k})+(-1)^{\nu}E_q({\bf k})]^2 
+ D^2({\bf k})}},\ \ \ \nu=1,2,
\label{eigen}
\end{eqnarray}
where
$
E_q({\bf k})=\sqrt{\epsilon^2({\bf k})+Q^2U^2}
$.

Minimizing the free energy function with respect to $Q$ and $D_0$,
and calculating the 
density of electrons $n$ led to the following 
mean-field equations
that describe the interplay between RAF and
SC for HTSCs with tetragonal symmetry:
\begin{eqnarray}
&&1 = {V\over4N}\sum_{{\bf k},\nu=1,2}{{(\cos k_x-\cos k_y)^2}\over{E_\nu}}
 \tanh({\beta E_\nu\over2}), \cr
&&1={U\over4N}\sum_{{\bf k},\nu=1,2}
(-1)^{\nu+1}{A_\nu\over E_q}\tanh({\beta E_\nu\over2}), \cr
&& n = -{1\over4N}\sum_{{\bf k},\nu=1,2}{A_\nu}\tanh({\beta E_\nu\over2})
+ {1\over2},
\label{gapeq3}
\end{eqnarray}
where 
\begin{eqnarray}
A_{\nu}({\bf k})=[{-\mu'({\bf k})-(-1)^\nu E_q({\bf k})}]/
{E_\nu({\bf k})}.
\label{anu}
\end{eqnarray}
In the case of crystals with orthorhombic symmetry it is possible that
$D_{\langle i,j\rangle_x}\ne-D_{\langle i,j\rangle_y}$
because the superconducting 
coupling constants $V_x$ along the $x$ axis and $V_y$ 
along the $y$ axis may differ. Then,
the superconducting gap takes on the form
$
D({\bf k})=\psi_s(\cos k_x + \cos k_y) + \psi_d(\cos k_x - \cos k_y)
$
with $\psi_s=V_xD_x-V_yD_y$, and $\psi_d=V_xD_x+V_yD_y$, which 
implies that it shows
$d$+$s$-pairing symmetry. This is
a consequence of the absence of invariance under
rotations by $\pi/2$ of the CuO$_2$ plane, and is therefore
consistent with arguments based on group theory
\cite{tsuei}.

\begin{figure}
\caption{The typical behavior of the rotating order parameter $Q$ 
and superconducting parameter $D_0$ with doping
is illustrated here for $U=2.8t$, $V=0.85t$, 
and $t'=-0.16t$. Temperature is $T=0.05t$. The behavior shown is
practically the same at zero temperature.
The doping values where $Q$ vanishes in both p and n-type
cases are interpreted as QCP.}
\begin{center}
\includegraphics[height=4.0cm]{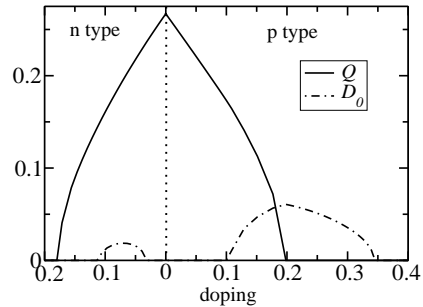}
\end{center}
\label{QD0vsdoping}
\end{figure}

\begin{figure}
\caption{The temperature dependence of $Q$ 
and $D_0$ is displayed for two
values of doping in the p-type case. 
The Hamiltonian parameters 
are $U=2.8t$, $V=0.85t$ and $t'=-0.16t$.}
\begin{center}
\includegraphics[height=4.0cm]{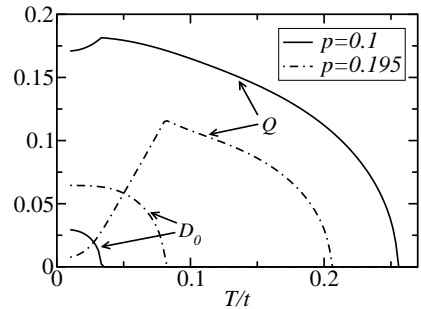}
\end{center}
\label{figQD0}
\end{figure}

\section{RAF and SC parameters versus temperature and doping}

\label{sec3}

The mean-field equations (\ref{gapeq3}) were solved numerically in order to 
get the parameters $Q$, $D_0$, and
average number of electrons per site and per spin
$n$. In hole-doped (p-type) systems, the density of holes
given by $p=1-2n$ is equivalent to 
the density of electrons missing below half-filling.
For electron-doped (n-type) systems
the density of electrons $n_e=2n-1$ is the density of electrons above
half-filling. Note that at half-filling $n=1/2$, so that $n_e=p=0$ 
in this case.

Fig. \ref{QD0vsdoping} shows RAF and SC parameters
$Q$ and $D_0$ versus doping for the Hamiltonian parameters
$U=2.8t$, $V=0.85t$, and $t'=-0.16t$. If RAF were not 
taken into account, SC would be optimum at half filling.
When RAF is allowed in, SC is destroyed near half filling, but coexists 
with RAF in the underdoped regime for p-type doping, and for all
doping values where $D_0\ne0$ in n-type case. The doping values where
$Q$ vanishes in both p  and n-type systems are identified as 
quantum critical points (QCP) \cite{azzouz2003}. 
In Fig. \ref{QD0vsdoping}, the QCP occurs within the superconducting
dome in the p-type system, but outside of the 
dome and deep in the overdoped regime for the n-type system. 
RAF's parameter
$Q$ has been proposed to model the PG in 
HTSCs \cite{azzouz2003,azzouz2003b},  and the PG temperature $T^*$ has been 
identified with the temperature below which $Q$ becomes nonzero.

\begin{figure}[t]
\caption{The temperature dependence of $Q$ 
and $D_0$ is displayed for an n-type case with $n_e=0.075$.
The Hamiltonian parameters 
used are $U=2.8t$, $V=0.85t$ and $t'=-0.16t$.}
\begin{center}
\includegraphics[height=4.0cm]{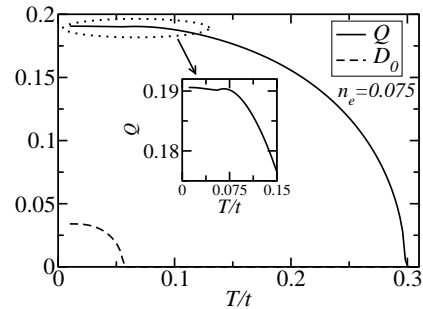}
\end{center}
\label{figQD0ntype}
\end{figure}

Figures \ref{figQD0} and \ref{figQD0ntype} display
the temperature dependence for 
$Q$ and $D_0$ for some given doping levels.
Using these kind of figures, the PG ($T^*$) and SC ($T_C$) 
temperatures were calculated in Ref. \cite{azzouz2003b} for 
p-type cuprates. Again the competition is apparent
between SC and RAF in the p-type case, because as soon as $D_0$
becomes nonzero RAF's order parameter $Q$ decreases monotonically
as seen in Fig. \ref{figQD0} for $p=0.1$ in the underdoped regime.
Note that the optimal doping for the Hamiltonian parameters used here
is $p\approx0.20$ in the p-type case, and $n_e\approx0.075$ 
in the n-type case; see Fig. \ref{QD0vsdoping}. For $p=0.195$ 
(close to optimal doping), it is interesting to note that 
$Q$ decreases significantly below $T_C$ when $D_0$ becomes nonzero. 
The behavior for n-type case is totally
different. For $n_e=0.075$ (optimal doping), $Q$ barely decreases
when $D_0$ becomes nonzero below $T_C$, then even increases slightly 
and saturates at low temperature as the inset of Fig. \ref{figQD0ntype}
shows.

The phase diagram obtained by letting in a 
Bardeen-Cooper-Schrieffer \cite{bcs} picture 
$T^*\sim Q(T=0)$ and $T_C\sim D_0(T=0)$
in Fig. \ref{QD0vsdoping} is in qualitative agreement with experiment
for p-type La$_{2-x}$Sr$_x$CuO$_4$ and n-type 
Nd$_{2-x}$Ce$_x$CuO$_4$ cuprates. For the latter, the PG is reported to vanish
almost when SC does in the overdoped regime \cite{helm2009}. For the former
significant experimental evidence suggests the disappearance of the PG
rather closer to the optimal doping \cite{tallon2001}.

%
%
 
In RAFT, the thermal average of the spin operators $S^x_i$ and $S^y_i$
in a frame rotating with angular frequency 
$\omega_{sf}$ with the rotating local magnetization are
$\langle S^x_i\rangle=(\langle S^+_i\rangle + 
\langle S^-_i\rangle)/2=(-1)^{x_i+y_i}|Q|$, 
and $\langle S^y_i\rangle=0$, respectively,
and 
by construction, the static magnetization along the quantization axis
$z$ is zero in order to satisfy the Mermin-Wagner theorem at finite temperature.
Here $x_i$ and $y_i$ are the $x$ and $y$ coordinates of site $i$.
RAF is predicted to exist in a purely two-dimensional electronic 
system or in a three-dimensional system of electrons
where either thermal fluctuations at high temperature
or doping even at lower temperature prevents three-dimensional long-range
N\'eel order from occurring. N\'eel order, 
which has not been taken into
account so far in RAFT, occurs below $T_N<T^*$ in the vicinity of
half-filling. As is well known, 
this antiferromagnetic phase 
consists of a static magnetization plus quantum 
spin waves, which exist for all allowed
wavevectors. How then does RAF evolve into 
static antiferromagnetism when
temperature is lowered below $T_N$ for a given doping density,
and how does static antiferromagnetism give way 
to the pseudogap phase when doping increases away from half-filling at 
a given temperature?
The key point in answering these questions may perhaps 
reside in the fact that
RAF has been interpreted as a single $q=(\pi,\pi)$ wave \cite{azzouz2012}.
We conjecture that
when temperature is lowered across $T_N$, the static magnetization sets in 
due to the three-dimensional coupling between the copper-oxygen layers. 
The establishment of three-dimensional
long-range order naturally 
allows other spin waves with $q\ne(\pi,\pi)$ to settle in along with 
the $q=(\pi,\pi)$ spin wave present in RAF, a mechanism which 
causes the loss of  RAF.
In this conjecture, the PG is a consequence of purely two-dimensional
physics, but the N\'eel order is as is well known due to three-dimensional
physics. In future investigations,
we plan to seek the mechanism for the phase change from 
N\'eel order to RAF, and vice versa.

%
%

\begin{figure}[t]
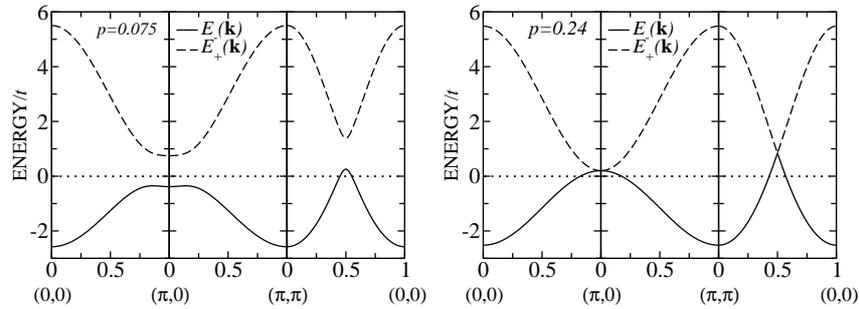

\caption{The energies $E_-({\bf k})$ and 
$E_+({\bf k})$ are plotted versus ${\bf k}$ 
along symmetry lines of the Brillouin zone. The Hamiltonian parameters
are $U=2.8t$, $V=0.85t$ and $t'=-0.16t$, and  
hole doping is $p=0.075$ in the underdoped phase 
for the figure on the left, and $p=0.24$ in 
the overdoped regime for the figure on the right. 
The dotted horizontal line indicates the position 
of the Fermi energy. Temperature is $T=0.1t$
and $D_0=0$.}
\begin{center}
\includegraphics[height=4.0cm]{./spectrump0.075.eps}\ \
\includegraphics[height=4.0cm]{./spectrump0.24.eps}
\end{center}
\label{spectrum p-type}
\end{figure}

\begin{figure}[t]
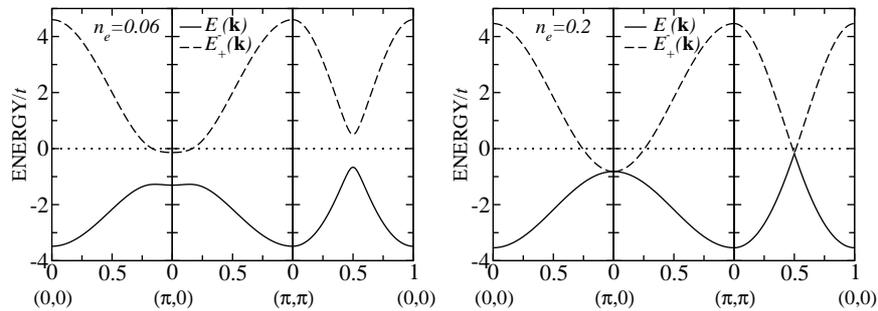

\caption{The energies $E_-({\bf k})$ and 
$E_+({\bf k})$ are plotted versus ${\bf k}$ 
along symmetry lines of the Brillouin zone. The Hamiltonian parameters
are $U=2.8t$, $V=0.85t$ and $t'=-0.16t$, and  
electron doping is $n_e=0.06$ in the underdoped phase 
for the figure on the left, and $n_e=0.2$ in 
the overdoped regime for the figure on the right. 
Temperature is $T=0.1t$
and $D_0=0$.}
\begin{center}
\includegraphics[height=4.0cm]{./spectrumn0.06.eps}\ \
\includegraphics[height=4.0cm]{./spectrumn0.2.eps}
\end{center}
\label{spectrum n-type}
\end{figure}

\begin{figure}[t]
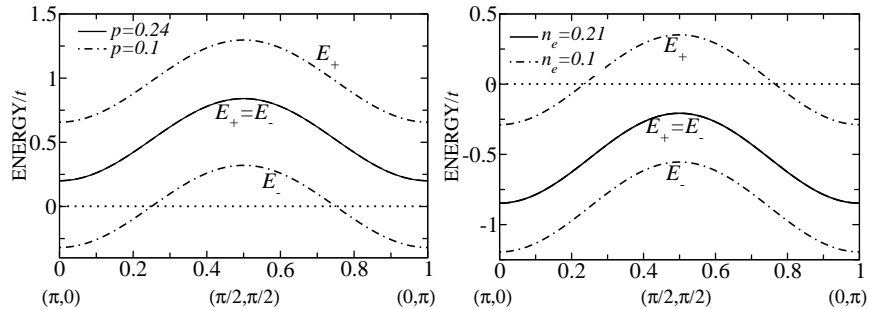

\caption{The energy spectra along the boundary of the magnetic 
Brillouin zone ($(\pi,0)\to (0,\pi)$) is shown. Temperature is $T=0.1t$
and $D_0=0$.}
\begin{center}
\includegraphics[height=4.0cm]{./spectrumkxpluskyPIp.eps}
\includegraphics[height=4.0cm]{./spectrumkxpluskyPIn.eps}
\end{center}
\label{spectrum RBZ side}
\end{figure}

\section{Doping dependence of electronic structure}

\label{sec4}

\subsection{Analysis of energy spectra}

As mentioned 
in the previous section,
the appearance of RAF below a critical value of doping as the latter is reduced
from overdoped to underdoped regime
for p-type or n-type systems at zero temperature
has been interpreted as a
QCP.
The case of p-type
has been discussed before \cite{azzouz2003,azzouz2003b,azzouz2005}.
This QCP induces a reconstruction of the FS practically 
in the same way an ordinary
spin-density order does \cite{chubukov1997}.
However, RAF is not an ordinary spin density order 
as explained in Section \ref{sec2}.
Figure \ref{spectrum p-type}
shows energy spectra along symmetry lines of the Brillouin zone for $T=0.1t$.
For $p=0.075$, one clearly sees a gap at $(\pi,0)$, in agreement 
with experimental data for La$_{2-x}$Sr$_x$CuO$_4$ \cite{yoshida2007}.
Also a small hole-like band is seen along the diagonal around 
$(\pi/2,\pi/2)$. The presence of the gap at $(\pi,0)$
for this doping and the small hole-like band
in the vicinity of $(\pi/2,\pi/2)$ are due to 
the nonzero value of 
RAF's order parameter $Q$. This gap is responsible 
for the PG behavior in the underdoped regime.
The hole-like band is also seen along 
the RBZ boundary [$(\pi,0)\to(0,\pi)$] as shown in 
Fig. \ref{spectrum RBZ side} for $p=0.1$.
For $p=0.24$ in the overdoped regime, the PG has closed and the 
hole-like pocket has reached the $(\pi,0)$ and $(0,\pi)$ points as can be seen
along the RBZ boundary in Fig. \ref{spectrum RBZ side}.
Along this boundary $E_+({\bf k})=E_-({\bf k})$, when $Q=0$ at $T=0.1t$,
is above the chemical potential all the way between
$(\pi,0)$ and $(0,\pi)$.

For the n-type case, Figure \ref{spectrum n-type} displays the spectra
for doping $n_e=0.06$ in the underdoped regime, and for $n_e=0.2$
well in the overdoped regime for $T=0.1t$.
The PG behavior is now a consequence 
of a gap at $(\pi/2,\pi/2)$, and a small
electron pocket forms near $(\pi,0)$. For $n_e=0.2$, the PG is zero because
$Q$ has vanished, and the electron pocket at $(\pi,0)$ joined
that at $(0,\pi)$. This can be understood by examining the spectrum along
the RBZ boundary which gives a completely full band along this direction.
For example, for $n_e=0.21$ in Fig. \ref{spectrum RBZ side}, 
$E_+({\bf k})=E_-({\bf k})<0$, which means that these bands are full.
The above analysis can be made even more transparent
by calculating the FS, a task 
undertaken below.

\subsection{Evolution of the Fermi surface with doping}

In RAFT, the occupation probability 
$n({\bf k})$ was defined by writing the average 
number of electron per spin
and site $n$ in equation (\ref{gapeq3}) as
%
%
%
%
%
$n\equiv {1\over N}\sum_{\bf k} n({\bf k})$, 
%
%
which yields \cite{azzouz2003}
\begin{eqnarray}
n({\bf k}) = -{1\over4}\sum_{\nu=1,2}{A_\nu({\bf k})}
\tanh[{\beta E_\nu({\bf k})\over2}]
+ {1\over2}.
\label{fermi2}
\end{eqnarray}
$A_\nu({\bf k})$ 
is given in Eq. (\ref{anu}).
$n({\bf k})$ was then interpreted as 
the probability that the state with wave vector
${\bf k}$ is occupied by an electron with spin up or down.

Ronning {\it et al.} \cite{ronning1998}
extracted $n({\bf k})$
by integrating ARPES energy distribution curves over energy
for the material Ca$_2$CuO$_2$Cl$_2$, then deduced the FS
by locating the steepest drops in $n({\bf k})$ 
in analogy with a Fermi gas. Also,
using the same method the Fermi surface
for Bi$_2$Sr$_2$CaCu$_2$O$_{8+\delta}$ in the overdoped regime was
obtained.
Here we implement 
the same argument in RAFT, namely the FS is determined by the sharp 
drops in the occupation probability.
This method was also used in Ref. \cite{azzouz2010} and gave results
in agreement with the determination of the FS using 
the spectral function.

\begin{figure}[t]
\caption{The occupation probability $n({\bf k})$
is shown in the Brillouin zone.
The hole densities are shown on the graphs.
The Hamiltonian parameters are $U=2.8t$, $V=0.85t$, 
and $t'=-0.16t$, and $T=0.1t$.}
\begin{center}
\includegraphics[height=7.0cm]{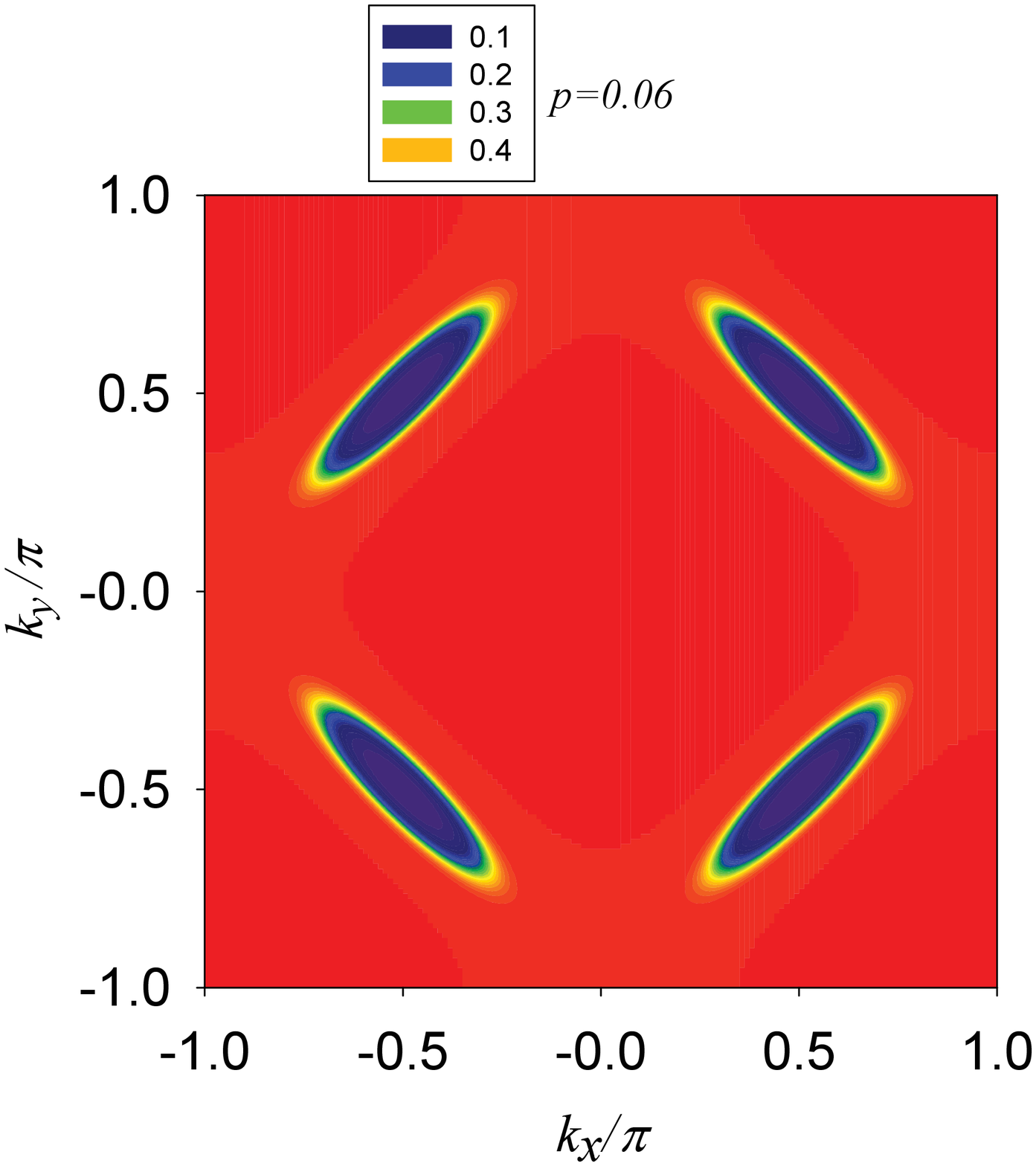}
\includegraphics[height=7.0cm]{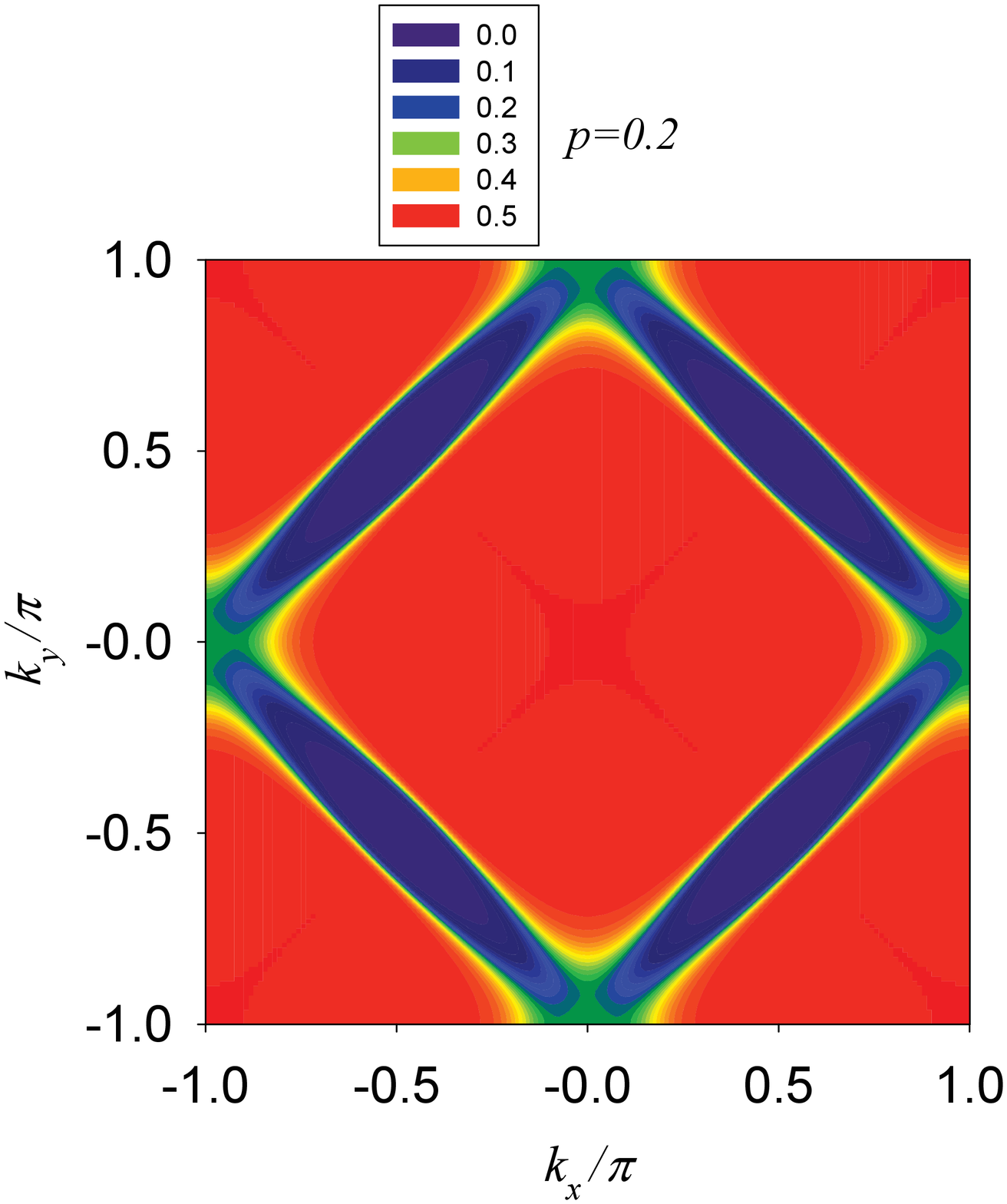}
\includegraphics[height=7.0cm]{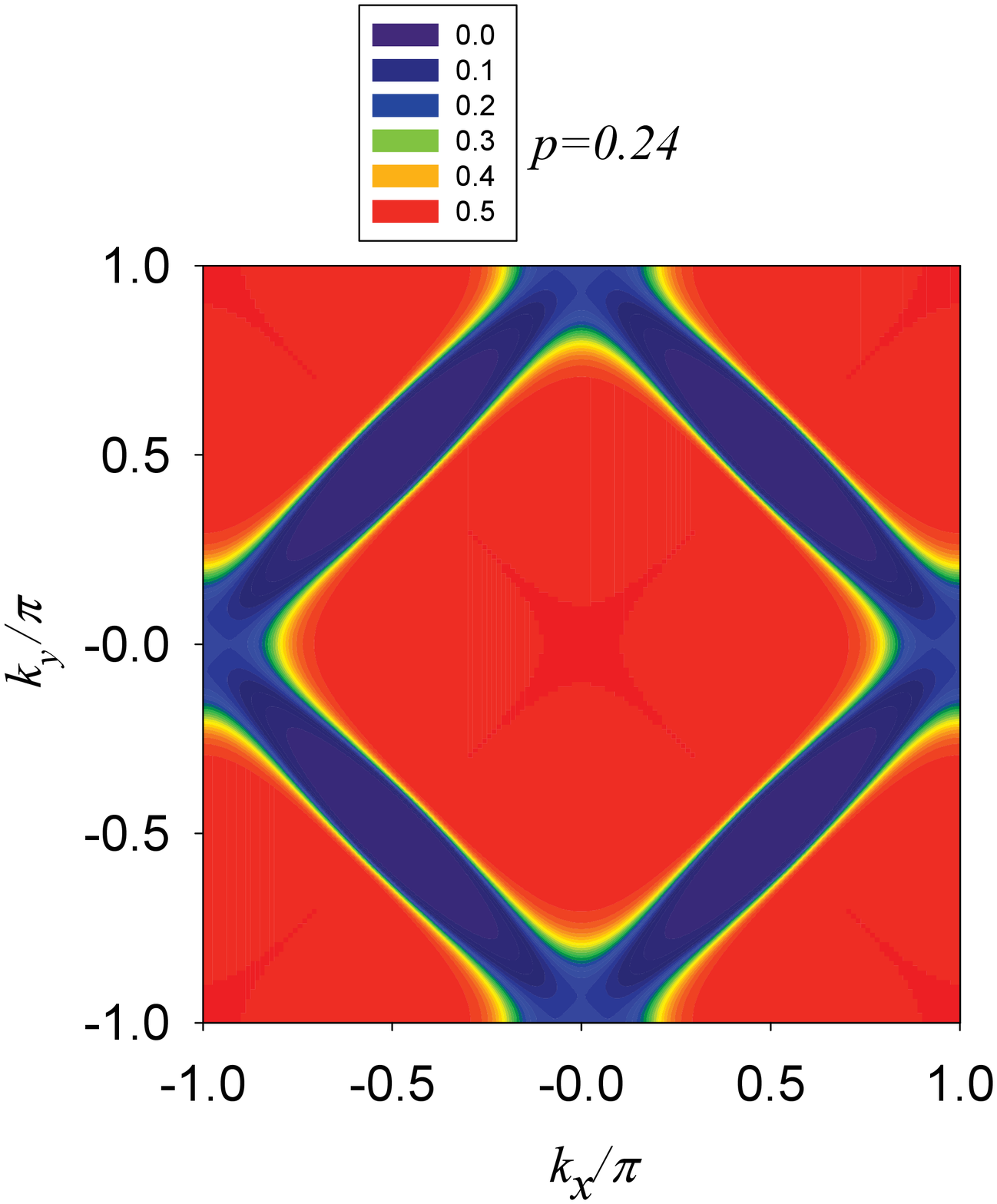}
\end{center}
\label{FS p-type}
\end{figure}

Fig. \ref{FS p-type} shows two-dimensional plots
of $n({\bf k})$ for three doping levels
in the p-type case for Hamiltonian parameters $U=2.8t$, $V=0.85t$, 
and $t'=-0.16t$ at temperature $T=0.1t$. This temperature is above
any transition temperature for SC.
The FS 
is made of hole pockets around $(\pm\pi/2,\pm\pi/2)$ 
in the underdoped regime as shown for $p=0.06$. 
The energy spectra in Figs. \ref{spectrum p-type} and \ref{spectrum RBZ side}
show well that in the presence 
of the PG, the upper $E_+$ and lower $E_-$ bands are separated by gaps
along all the symmetry lines in the underdoped regime. Hole-like pockets
can clearly be seen for $p=0.075$ in Fig. \ref{spectrum p-type} around 
$(\pi/2,\pi/2)$.  
Around optimal doping $p=0.2$, the hole pockets
reach the points $(\pm\pi,0)$ and $(0,\pm\pi)$. In the overdoped regime,
where the PG is zero, the FS is made of large contours around 
$(0,0)$ and $(\pi,\pi)$
as can be seen in Fig. \ref{spectrum p-type} for $p=0.24$. For the latter,
because the PG is zero the upper band $E_+$ and lower band $E_-$ touch 
at $(\pm\pi,0)$, $(0,\pm\pi)$ and $(\pm\pi/2,\pi/2)$ to form 
a tight-binding spectrum given by 
$E_\pm({\bf k})=\pm|2t(\cos k_x +\cos k_y)|
- 4t'\cos k_x\cos k_y -\mu+Un$. 
The presence of the absolute value in this tight-binding 
energy is a consequence of the limit $Q\to0$ 
in $\sqrt{\epsilon^2({\bf k})+U^2Q^2}$ for the overdoped regime.
In the p-type case,
the FS in RAFT thus evolves strongly 
with doping. It reconstructs at the QCP doping where $Q$ vanishes.
Its topology changes from small hole-like pockets in 
the underdoped regime below this QCP
to large contours in the overdoped regime.
This is qualitatively consistent with 
the quantum oscillations observed in resistivity
by Doiron-Leyraud {\it et al.} \cite{doiron2007}
for YBa$_2$Cu$_3$O$_{6.5}$, which indicated
that a well defined small FS characterizes this underdoped cuprate.
Subsequent work by Sebastian {\it et al.} \cite{sebastian2010} 
for YBa$_2$Cu$_3$O$_{6+x}$ supported the existence of small
closed pockets in the underdoped regime as well.

\begin{figure}[t]
\caption{The occupation probability $n({\bf k})$
is shown in the Brillouin zone.
The electron densities above half-filling $n_e$ are shown on the graphs.
The Hamiltonian parameters are $U=2.8t$, $V=0.85t$ 
and $t'=-0.16t$ and temperature is $T=0.1t$.}
\begin{center}
\includegraphics[height=7.0cm]{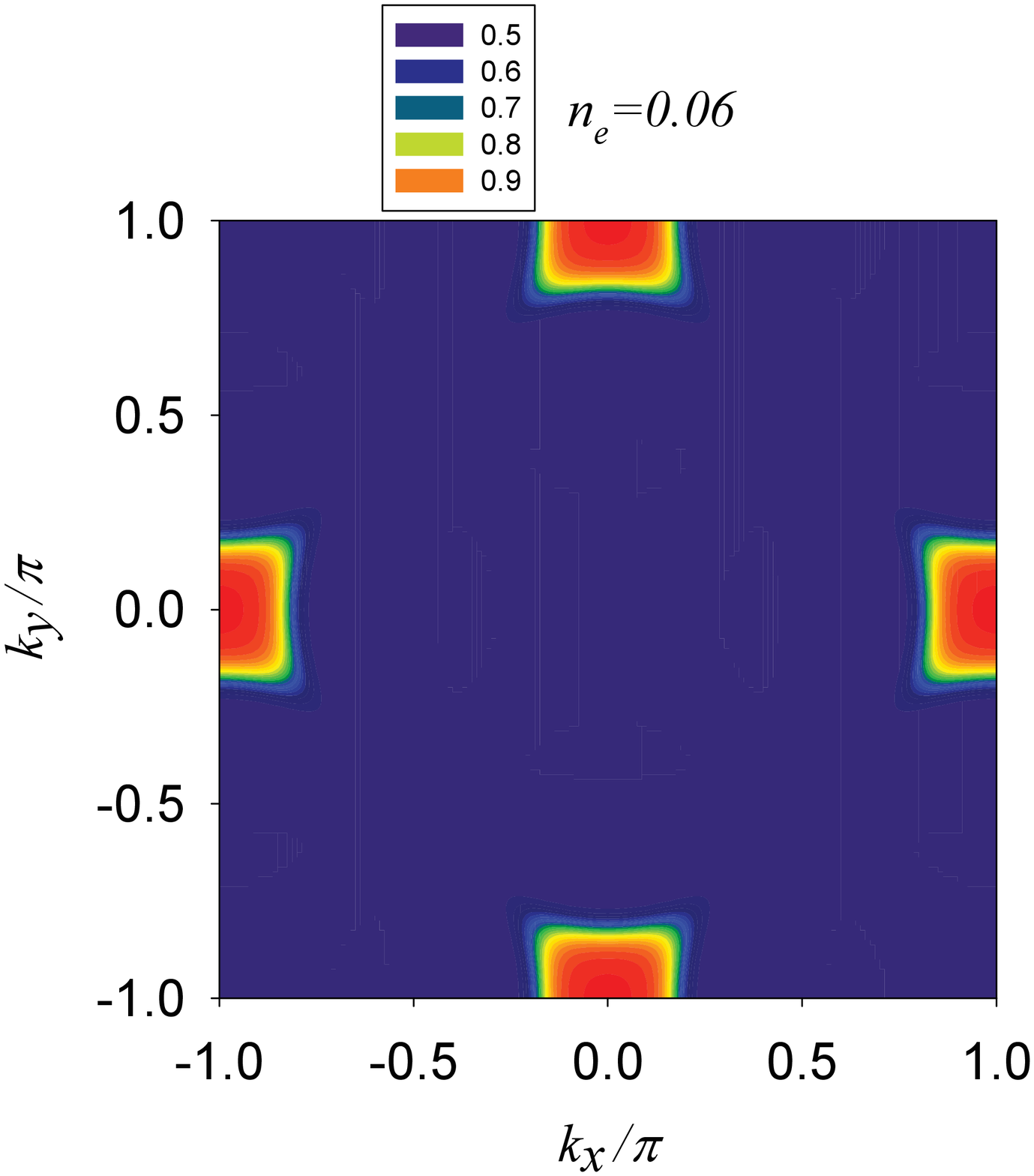}
\includegraphics[height=7.0cm]{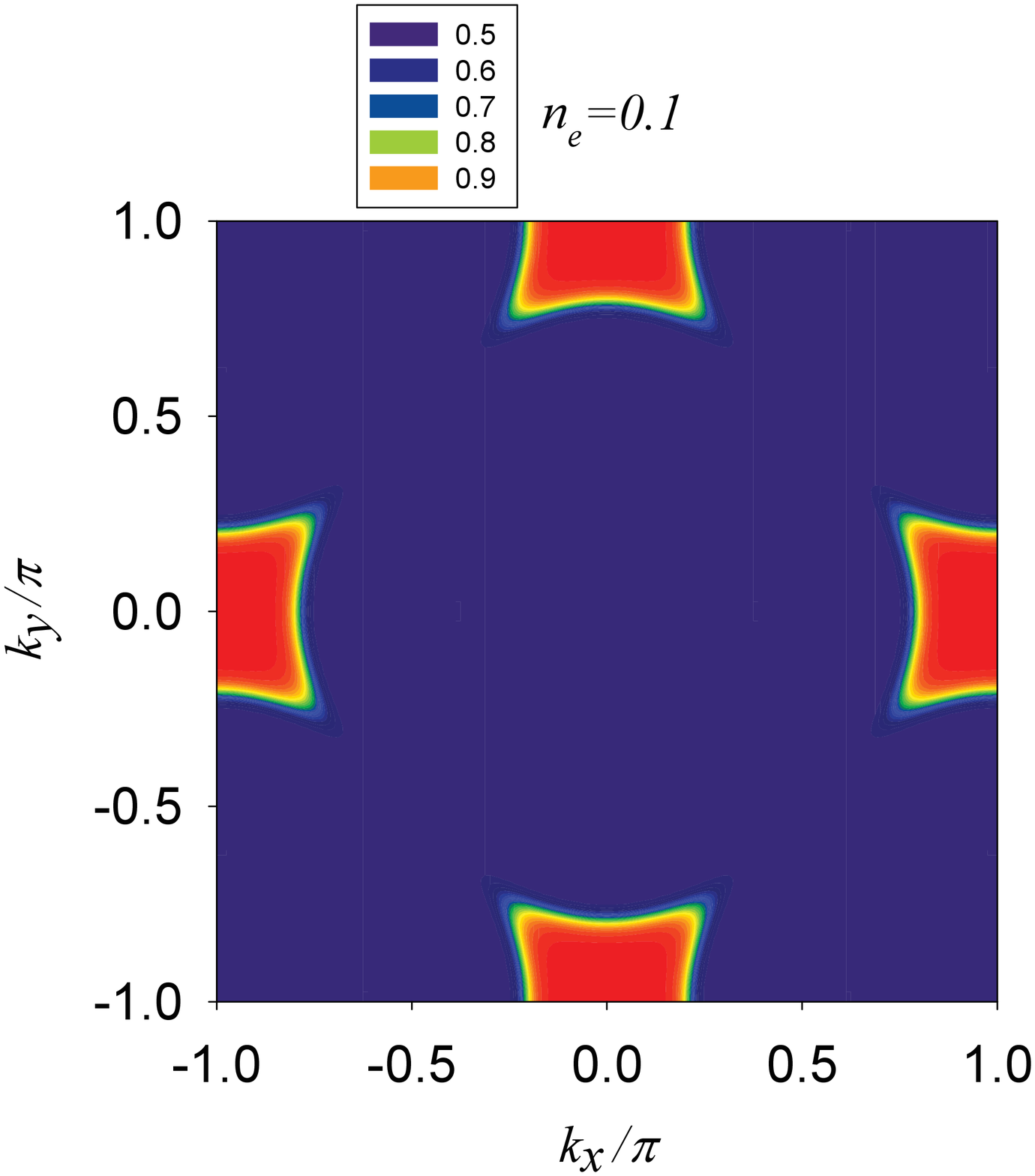}
\includegraphics[height=7.0cm]{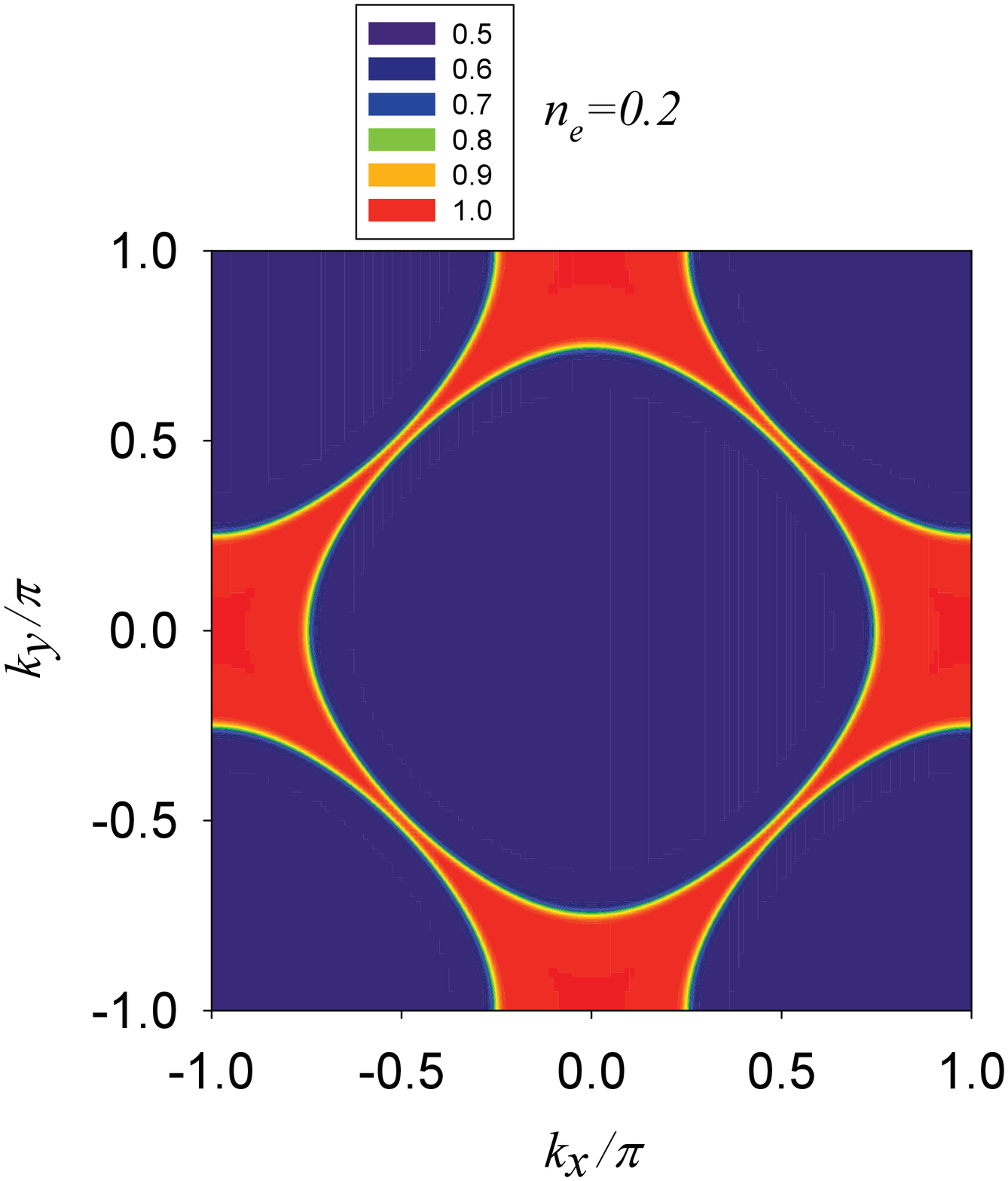}
\end{center}
\label{FS n-type}
\end{figure}

The calculated FS undergoes also a significant reconstruction when doping 
changes from p-type to n-type across half-filling,
(note that RAFT is only valid outside of the HTSCs' AF phase around half-filling).
In RAFT, for the Hamiltonian parameters considered here
the FS in the underdoped regime for n-type cuprates  
consists of electron pockets around points $(\pm\pi,0)$ and 
$(0,\pm\pi)$, rather than pockets around $(\pm\pi/2,\pm\pi/2)$ in 
the underdoped regime of p-type cuprates.
This is clearly seen in Fig. \ref{FS n-type} for $n_e=0.06$ and $n_e=0.1$,
and is consistent with 
the energy spectra in Fig. \ref{spectrum n-type}, which
show well the existence 
of a small electron pocket at $(\pi,0)$ for $n_e=0.06$.
When the PG vanishes in the overdoped regime, the electron pockets join 
to form large contours as seen for $n_e=0.2$. Armitage 
{\it et al.} \cite{armitage2002} reported ARPES data for n-type material
Nd$_{2-x}$Ce$_x$CuO$_4$ which can be interpreted as revealing 
the existence of 
pockets around $(\pi,0)$ and symmetric points
in the underdoped regime. Also, Matsui {\it et al}. \cite{matsui2007}
measured the evolution of the FS with doping for this material
using ARPES. A close look at Fig. 1 of their work reveals
a FS mainly near {\bf k}-points $(\pi,0)$ and
$(0,\pi)$ for doping $x=0.13$, but the FS evolves 
into larger contours joining these two points for the larger
doping levels $x=0.16$ and $x=0.17$. 
Note that if one symmetrizes Matsui {\it et al.}'s FSs
about the line joining $(\pi,0)$ and $(0,\pi)$, one will get
FSs that look similar to those calculated here, 
and shown in Fig. \ref{FS n-type}. 
Their measurements were done only along
the FS in the tight-binding limit. 
We predict that if measurements were performed along 
the image of this tight-binding 
FS with respect to line $(\pi,0)$-$(0,\pi)$, 
then one would obtain a FS that looks like ours. 
Also, the reconstruction of the FS, as illustrated 
in Fig. \ref{FS n-type}, at
the QCP doping where $Q$ vanishes (so where the PG vanishes) is in  
agreement with SdH oscillation results of 
Helm {\it et al.} \cite{helm2009}
for the above material. SdH oscillations 
revealed a FS evolving from small
pockets to large contours as doping 
goes from underdoped regime 
to overdoped regime.

\section{Conclusions and Discussion}

\label{conclusions}

We studied the reconstruction of the Fermi surface
under the effect of hidden rotating antiferromagnetic
order in both p-type and n-type high-$T_C$ cuprates. For the Hamiltonian
parameters used here,
the Fermi surface
in p-type cuprates reconstructs due to 
rotating antiferromagnetism at the quantum critical point
near optimal doping where the pseudogap vanishes. This Fermi surface 
consists of hole pockets around
$(\pm\pi/2,\pm\pi/2)$
in the underdoped regime, but changes to large closed contours
in the overdoped regime.
For n-type cuprates, the location and topology of the Fermi surface 
is different than in p-type materials.  
The hole pockets are 
of a form resembling squares around 
$(\pm\pi,0)$ and $(0,\pm\pi)$ in deep underdoped n-type systems. 
When the pseudogap becomes zero beyond the quantum critical point 
in the overdoped regime, the 
Fermi surface changes to large closed contours. These results 
are in good qualitative agreement with experimental data
for La$_{2-x}$Sr$_x$CuO$_4$ and Nd$_{2-x}$Ce$_x$CuO$_4$ 
for the set of Hamiltonian
parameters used in the present calculations.

The issue of the applicability of the
rotating antiferromagnetism theory
for the high-$T_C$ materials ought to 
be discussed in the context of other
experimental results.
For example, one needs to analyze in this theory
the unusual antiferromagnetic order observed using polarized 
neutron scattering by Fauqu\'e {\it et al.} \cite{fauque2006} in YBa$_2$C$_3$O$_{6+\delta}$,
and by Li {\it et al.} \cite{li2008} in HgBa$_2$CuO$_{4+\delta}$. 
These polarized neutron measurements
probed the spin-flip response. While it is not yet
clear how
to interpret these measurements in the framework of 
RAFT, it is
interesting to note that rotating 
antiferromagnetism is also based on the 
spin-flip processes in the Hubbard model
\cite{azzouz2003,azzouz2003b}. So,
could the occurrence of long-range coherence for
these spin flip processes as discussed in Ref. \cite{azzouz2012}
yield a magnetic signal identical to 
that observed using polarized neutrons? 
This question needs 
to be addressed both experimentally and theoretically.
One also needs to reconcile the
rotating antiferromagnetism theory with other 
observed types of orders like the 
charge-density wave order observed by Chang {\it et al.} 
\cite{chang2012} found using x-ray diffraction, 
and with the observation by Shekhter {\it et al.} 
\cite{shekhter2012} of a second-order phase transition
in resonant ultrasound spectroscopy. 
The charge-density wave order has been observed 
well below the pseudogap temperature 
$T^*$; it can thus not be claimed to be responsible for 
the pseudogap state in any way, and such an order
is not included in RAFT.
Regarding the observation 
at $T^*$ of a second-order phase transition 
in resonant ultrasound spectroscopy,
we stress that even though RAF is a dynamic order,
the rotating order parameter has a magnitude
that behaves as in a second-order phase transition
\cite{azzouz2003,azzouz2003b}, thus
in qualitative agreement with this experimental finding.
Finally, Dean {\it et al.} \cite{dean2012} reported spin-wave (magnon)
like excitations in a single layer La$_2$CuO$_4$ that resemble the
magnon excitations observed in the bulk material La$_2$CuO$_4$.
As long-range antiferromagnetic order 
is ruled out in a single layer at finite temperature
due to thermal 
spin fluctuations, one needs to find an explanation for this result
outside of the linear spin-wave theory. Also, Dean {\it et al.}
ruled out the possibility of interpreting
their finding within the resonating-valence bond theory.
The question that naturally arises is whether 
fluctuations beyond the mean-field point in the 
rotating antiferromagnetism theory
can mimic this observed magnon-like dispersion. 
The calculation of the dispersion due to such fluctuations
in underway and will be reported on in the future.

%
%
%

%
%

%
%
%
%
%
%
%



%
%
%
%
%

\bibliographystyle{mdpi}

\end{document}